\newif\ifsubmission
\submissionfalse

\documentclass{bmvc2k}

\ifsubmission
  \bmvcreviewcopy{??}
\fi

\usepackage{amssymb,amsfonts}
\usepackage{booktabs}
\usepackage{multirow}
\usepackage{algorithm}
\usepackage{algorithmic}
\usepackage{enumitem}
\usepackage{xcolor}

\newcommand{\method}{KNT}

\def\eg{\emph{e.g}\bmvaOneDot}

\title{Keyed Nonlinear Transform: Lightweight Privacy-Enhancing Feature Sharing for Medical Image Analysis}

\ifsubmission
  \addauthor{Anonymous}{}{1}
  \addauthor{Anonymous}{}{1}
  \addinstitution{Anonymous Institution}
  \runninghead{Anonymous}{Keyed Nonlinear Transform}
\else
  \addauthor{Haebom Lee}{hblee@oolu.team}{1}
  \addauthor{Gyeongjung Kim}{gjkim@oolu.team}{1}
  \addinstitution{OOLU Soft Co., Ltd.}
  \runninghead{Lee and Kim}{Keyed Nonlinear Transform}
\fi

\begin{document}
\maketitle

\begin{abstract}
Feature sharing via split inference offers a lightweight alternative to federated learning for resource-constrained hospitals, but transmitted features still leak patient identity information and lack practical mechanisms for controlled feature sharing. We propose Keyed Nonlinear Transform (KNT), a drop-in feature transformation that applies key-conditioned obfuscation to intermediate representations. KNT reduces re-identification AUC from 0.635 to 0.586, corresponding to a $36\,\%$ reduction in above-chance identity signal, while introducing only $0.15\,ms$ CPU overhead, without backbone retraining, and preserving classification performance within $1.0\,pp$. Our analysis shows that KNT’s nonlinear transform prevents closed-form inversion and shifts recovery to iterative gradient-based optimization under full key compromise, substantially increasing inversion difficulty. The same transform generalizes to dense prediction tasks, incurring only a 4.4 pp Dice reduction on skin-lesion segmentation without retraining. These results position KNT as a practical and efficient privacy layer for split inference deployments.
\end{abstract}

\section{Introduction}
\label{sec:intro}

Medical images contain biometric information that can identify patients.
\citet{packhaeuser2022reid} demonstrated that chest X-rays enable patient re-identification with a verification AUC of 0.994, and subsequent work confirmed this threat at scale across over one million radiographs~\citep{macpherson2023reid}.
Hospitals need cloud-hosted deep learning for diagnosis, but transmitting raw images risks patient privacy.
Federated learning keeps data local but requires each institution to maintain capital-intensive GPU clusters, participate in synchronous training rounds, and manage model aggregation, overhead that is prohibitive for many resource-constrained hospitals.

Feature sharing via split inference~\citep{vepakomma2018split} is a lighter alternative: the hospital runs only a frozen feature extractor, and heavy computation happens on the cloud.
Image reconstruction from deep features is already difficult (\citet{xu2024fora} report SSIM 0.03 at layer 4 using a gradient-based attack without training data), so pixel-level image privacy is largely preserved by depth alone.
Patient re-identification, however, is a separate threat: identity can be inferred from feature similarity without reconstructing any image.
We confirm empirically that frozen ResNet-18 layer-4 features enable patient re-identification with AUC 0.635 (Section~\ref{sec:main}).
Without any feature protection mechanism, intercepted representations remain vulnerable to identity inference attacks.

Existing defenses fall into two categories, each with a critical limitation.
\emph{Keyless defenses} such as NoPeek~\citep{vepakomma2020nopeek}, Shredder~\citep{mireshghallah2020shredder}, and DISCO~\citep{singh2021disco} modify features without a secret, so anyone who observes the defense mechanism can reverse or circumvent it.
\emph{Keyed linear defenses} such as PPCMI-SF~\citep{bello2026ppcmisf} use secret-key-dependent transforms, but restrict themselves to orthogonal (linear) mixing, which is vulnerable to closed-form inversion~\citep{shi2025scalemia}.

We propose \emph{Keyed Nonlinear Transform} (\method{}), which combines three lightweight feature-protection components:
(1)~key-dependent spatial permutation of feature map positions,
(2)~per-patch multi-layer nonlinear transform with key-derived parameters, and
(3)~optional dimensionality reduction via projection ($C \to d$).
Spatial permutation disrupts spatial coherence, the multi-layer nonlinear transform prevents closed-form inversion, and projection (when $d < C$) reduces the transmitted information.
We position \method{} as a practical protection layer for split inference, and our adversarial analysis characterizes the boundaries of its protection.

Our contributions are:
\begin{enumerate}[leftmargin=*,itemsep=2pt]
  \item \textbf{A practical, drop-in feature transform.} \method{} is, to our knowledge, the first keyed nonlinear defense applied post-hoc to split inference features. Unlike keyed chaotic dynamics~\citep{fagan2025chaotic}, which modifies inference weights, \method{} transforms transmitted features directly with negligible overhead (Table~\ref{tab:generalization}), no backbone retraining, and classification utility within 1.0\,pp at $d{=}512$.

  \item \textbf{Privacy-utility evaluation across tasks and backbones.} We evaluate image reconstruction (SSIM), identity signal reduction (verification AUC), and downstream performance on both classification (five MedMNIST datasets) and dense prediction (ISIC 2018 skin-lesion segmentation), using two backbones and multiple projection dimensions.

  \item \textbf{Transparent adversarial analysis.} We evaluate key compromise under both analytic and gradient-based attacks. Feature-level recovery is possible under key compromise, but image reconstruction from deep features remains extremely difficult with current methods~\citep{xu2024fora}. We discuss deployment guidance including key management and dimensionality reduction as a secondary defense.
\end{enumerate}

\section{Related Work}
\label{sec:related}

\paragraph{Patient re-identification from medical images.}
Chest X-rays contain biometric features (bone structure, body habitus, cardiac silhouette) that enable patient re-identification even after standard anonymization.
\citet{packhaeuser2022reid} achieved verification AUC~0.994 on ChestX-ray14 using a trained verification network, and \citet{macpherson2023reid} extended this to over one million radiographs with triplet-loss metric learning (Precision@1 = 0.976), confirming the threat at scale.
Image-space mitigations such as PriCheXy-Net~\citep{packhaeuser2023prichexy} reduce re-identification AUC from 0.818 to 0.577 via learned deformation fields, but they operate on raw images and do not address the privacy of features transmitted in split inference.

\paragraph{Feature-level attacks on split learning.}
FORA~\citep{xu2024fora} demonstrated feature-oriented reconstruction attacks on split learning, achieving SSIM~0.83 from shallow features (layer~2) but only SSIM~0.03 from deep features (layer~4); this depth-induced robustness applies to pixel-level reconstruction only and does not block identity inference via feature similarity, which we evaluate separately as verification AUC.
Scale-MIA~\citep{shi2025scalemia} showed that linear layers are analytically invertible, enabling scalable model inversion.
\citet{erdogan2022unsplit} showed that model-inversion and label-inference attacks succeed even without access to training data, and \citet{gao2023pcat} demonstrated that a pseudo-client can steal data and functionality from split learning using minimal public data.
\citet{higgins2025destabilizing} systematically evaluated existing defenses, finding that most provide limited protection.

\paragraph{Keyless defenses.}
NoPeek~\citep{vepakomma2020nopeek} minimizes distance correlation between inputs and activations.
Shredder~\citep{mireshghallah2020shredder} learns noise distributions; DISCO~\citep{singh2021disco} zeros sensitive channels.
All three are keyless and require training-time modification.

\paragraph{Differential privacy for feature release.}
Differential privacy~\citep{dwork2014algorithmic} provides formal $(\varepsilon, \delta)$ guarantees against statistical inference. DP-SGD~\citep{abadi2016deep} adds calibrated Gaussian noise to gradients during training; for inference-time release, the Gaussian mechanism perturbs released features with $\mathcal{N}(0, \sigma^2 I)$ calibrated to the $L_2$ sensitivity, and has been applied to medical-imaging deep learning~\citep{ziller2021medical}. We compare against this formal baseline in Section~\ref{sec:analysis}.

\paragraph{Keyed and spatial defenses.}
\citet{yao2022patchshuffling} proposed patch shuffling for Vision Transformers, permuting patch tokens to disrupt spatial coherence.
\citet{niwa2024secretkey} applied secret-key transforms to speech CNNs.
PPCMI-SF~\citep{bello2026ppcmisf} uses key-dependent orthogonal mixing on autoencoder latent features for medical segmentation.
\citet{fagan2025chaotic} proposed keyed chaotic dynamics for privacy-preserving inference.
LoFt~\citep{yu2024loft} showed that low-rank filtering reduces privacy leakage, supporting dimensionality reduction as a privacy mechanism.

\method{} differs from prior work by combining key-conditioned feature protection, nonlinear transforms, and spatial disruption in a single post-hoc transform.
Whereas NoPeek/Shredder/DISCO operate without a secret and PPCMI-SF restricts itself to orthogonal (linear) mixing, \method{} pairs a secret key with nonlinear transforms that resist closed-form algebraic inversion.
InstaHide~\citep{huang2020instahide} was broken by \citet{carlini2021instahide} because its linear mixing is analytically invertible; \method{} applies element-wise ReLU, which prevents closed-form inversion, though gradient-based optimization can still recover features under key compromise (Section~\ref{sec:keycompromise}).
Recent theoretical work by \citet{xiao2024formal} proved that any utility-preserving encoding necessarily leaks information, placing a fundamental limit on all lightweight privacy transforms including \method{}.

\section{Method}
\label{sec:method}

Given spatial features from a frozen backbone (Section~\ref{sec:features}), \method{} transforms them into key-protected representations through three sequential components: spatial permutation (Section~\ref{sec:perm}), per-patch keyed nonlinear transform (Section~\ref{sec:knt}), and optional dimensionality reduction (Section~\ref{sec:proj}).

\subsection{Feature Extraction}
\label{sec:features}

Given an input image $\mathbf{x} \in \mathbb{R}^{H_0 \times W_0 \times 3}$, the client computes features using a frozen ResNet-18 backbone truncated after layer~4:
\begin{equation}
  \mathbf{F} = \phi(\mathbf{x}) \in \mathbb{R}^{H \times W \times C},
\end{equation}
where $H = W = 7$ and $C = 512$ for $224 \times 224$ input images.
The backbone is frozen (no fine-tuning), so the feature extractor is deterministic and identical across institutions.
While we use ResNet-18 as the primary backbone throughout this paper, \method{} operates on any spatial feature map $H \times W \times C$ and is not tied to a specific architecture (Section~\ref{sec:analysis}).

\subsection{Spatial Permutation}
\label{sec:perm}

A key-derived permutation $\pi_k : \{1, \ldots, HW\} \to \{1, \ldots, HW\}$ shuffles the $HW = 49$ spatial positions of the feature map:
\begin{equation}
  \mathbf{F}^{\pi}_{i} = \mathbf{F}_{\pi_k(i)}, \quad i = 1, \ldots, HW.
\end{equation}
The permutation is generated deterministically from the key~$k$ using a seeded pseudorandom number generator (PRNG).
Spatial permutation breaks the spatial coherence of the feature map, disrupting attacks that exploit spatial structure (\eg, U-Net decoders that assume spatial correspondence).
Because downstream classification uses global average pooling over spatial positions, permutation has zero effect on classification accuracy (Table~\ref{tab:main}: \emph{Spatial permutation} achieves identical AUC to \emph{Raw features}).
For dense-prediction tasks where the decoder depends on the original spatial layout, the inverse permutation must be applied before decoding; this step can only be performed by a party holding the key (Section~\ref{sec:setup}).

\subsection{Per-Patch Keyed Nonlinear Transform}
\label{sec:knt}

At each spatial position~$i$, we apply an $L$-layer nonlinear transform with key-derived parameters:
\begin{equation}
  \mathbf{h}^{(0)}_i = \mathbf{f}^{\pi}_i, \quad
  \mathbf{h}^{(\ell)}_i = \text{ReLU}\!\bigl(\mathbf{W}^{(\ell)}_k \mathbf{h}^{(\ell-1)}_i + \mathbf{b}^{(\ell)}_k\bigr), \quad
  \mathbf{g}_i = \mathbf{h}^{(L)}_i,
  \label{eq:knt}
\end{equation}
where $\ell = 1, \ldots, L$.
All weight matrices $\mathbf{W}^{(\ell)}_k \in \mathbb{R}^{d \times d}$ and bias vectors $\mathbf{b}^{(\ell)}_k \in \mathbb{R}^{d}$ are drawn i.i.d.\ from $\mathcal{N}(0, 1/\sqrt{d})$ using the key~$k$ as the PRNG seed (the first layer is $\mathbf{W}^{(1)}_k \in \mathbb{R}^{d \times C}$ when $d \neq C$).
We use $L{=}2$ and $d{=}512$ as defaults in this paper.
The same parameters are applied at every spatial position.

Each ReLU layer zeroes components with negative pre-activations, preventing exact analytic inversion.
This distinguishes \method{} from linear keyed transforms~\citep{bello2026ppcmisf}: a linear transform $\mathbf{W}_k \mathbf{f}$ can be inverted given $\mathbf{W}_k$~\citep{shi2025scalemia}, but ReLU breaks the linear relationship needed for closed-form recovery.
Section~\ref{sec:keycompromise} analyzes residual attack surfaces under key compromise.

\subsection{Dimensionality Reduction}
\label{sec:proj}

The first-layer weight matrix $\mathbf{W}^{(1)}_k \in \mathbb{R}^{d \times C}$ with $C = 512$ simultaneously performs the nonlinear transform and optional dimensionality reduction.
The projection dimension~$d$ is a tunable deployment parameter controlling the privacy-utility-bandwidth tradeoff.
When $d = C$, $\mathbf{W}^{(1)}_k$ is square and no compression occurs, so privacy protection comes solely from the multi-layer nonlinear transform and spatial permutation.
When $d < C$, the first layer performs projection, providing additional privacy through information compression: at $d = 128$, $4\times$ compression reduces transmitted data from $25{,}088$ to $6{,}272$ values per image, while $d = 256$ offers an intermediate tradeoff (Section~\ref{sec:analysis}).

\subsection{Full \method{} Pipeline}
\label{sec:pipeline}

\begin{figure}[!htbp]
  \centering
  \includegraphics[width=\linewidth]{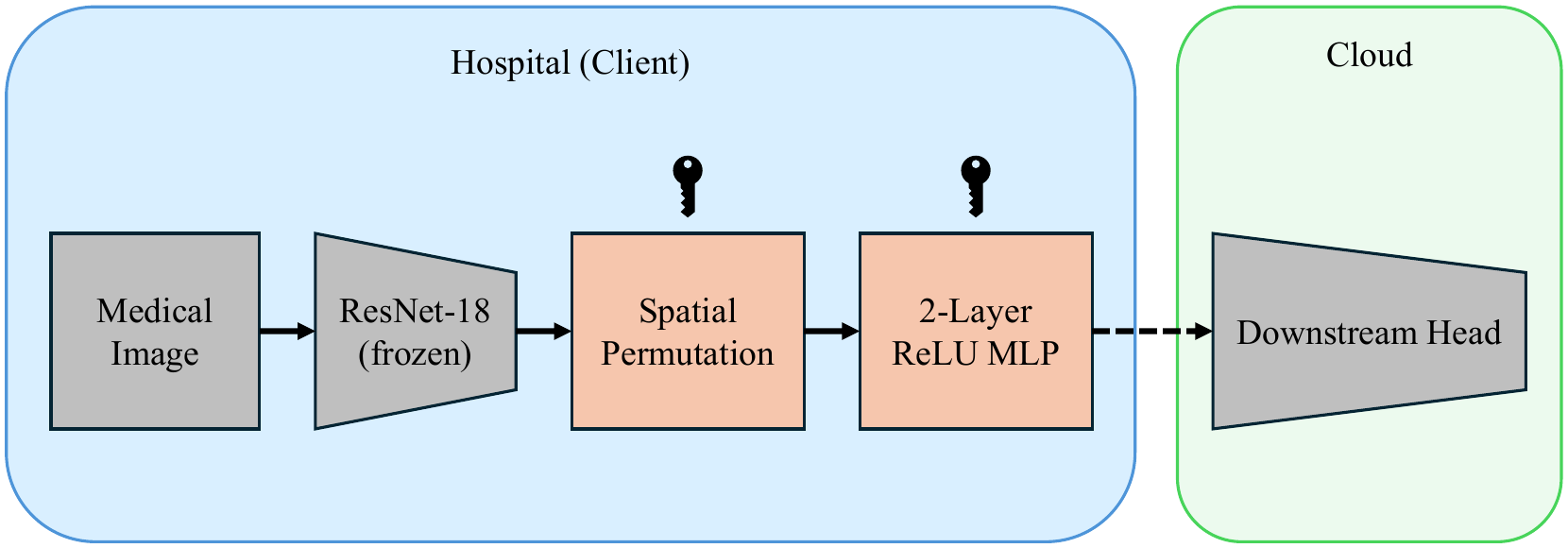}
  \caption{\textbf{Overview of the \method{} pipeline.} A frozen backbone extracts spatial features ($H \times W \times C$). The client applies key-derived spatial permutation followed by a 2-layer keyed nonlinear transform (Eq.~\ref{eq:knt}), with optional dimensionality reduction ($C \to d$). The transformed features are sent to the cloud server for downstream tasks (\eg, classification, segmentation).}
  \label{fig:pipeline}
\end{figure}

The complete transform, illustrated in Figure~\ref{fig:pipeline}, applies the three components sequentially:
\begin{equation}
  \mathbf{G} = \method{}(\mathbf{F}, k) = \text{ReLU}\!\bigl(\mathbf{W}^{(2)}_k \text{ReLU}(\mathbf{W}^{(1)}_k \cdot \text{Perm}_k(\mathbf{F}) + \mathbf{b}^{(1)}_k) + \mathbf{b}^{(2)}_k\bigr),
\end{equation}
where $\text{Perm}_k$ applies key-derived spatial permutation and the first layer optionally includes $512 \to d$ projection.
The output $\mathbf{G} \in \mathbb{R}^{7 \times 7 \times d}$ is transmitted to the server.

The server receives $\mathbf{G}$ and performs classification using a linear probe trained on \method{}-transformed features: global average pooling followed by a linear classifier.

\section{Experiments}
\label{sec:experiments}

\subsection{Experimental Setup}
\label{sec:setup}

\paragraph{Datasets.}
For classification, we use five datasets from MedMNIST v2~\citep{medmnistv2}.
\textbf{ChestMNIST} (112,120 chest X-rays, 14 binary labels, $224 \times 224$) is our primary evaluation dataset due to its established re-identification risk~\citep{packhaeuser2022reid}.
\textbf{DermaMNIST} (10,015 dermatoscopic images, 7 classes), \textbf{RetinaMNIST} (1,600 fundus photographs, 5 ordinal classes), \textbf{PathMNIST} (107,180 colorectal histology patches, 9 classes), and \textbf{PneumoniaMNIST} (5,856 pediatric chest X-rays, binary) test generalization across imaging modalities and task complexities.
For dense prediction, we use \textbf{ISIC 2018 Task~1}~\citep{codella2019isic2018} (2594 train / 100 val / 1000 test skin-lesion segmentation masks at $224 \times 224$).
All MedMNIST datasets follow the official train/validation/test splits provided by~\citet{medmnistv2}; for ChestMNIST, this corresponds to 78{,}468 / 11{,}219 / 22{,}433 patient-disjoint images.

\paragraph{Backbones and task heads.}
All features are extracted from a frozen ImageNet-pretrained ResNet-18 at the output of layer~4, yielding $7 \times 7 \times 512$ spatial feature maps (no fine-tuning); we additionally evaluate ResNet-50 for backbone generalization.
For classification, a linear probe (global average pooling $\to$ linear classifier $\mathbb{R}^d \to \mathbb{R}^K$) is trained on the (possibly transformed) features over 3 random seeds ($\pm$\,std).
AUC conventions follow each task: per-label binary AUC macro-averaged over 14 pathologies for multi-label ChestMNIST; one-vs-rest macro-averaged AUC for multi-class DermaMNIST / PathMNIST / RetinaMNIST; standard binary AUC for PneumoniaMNIST.
For dense prediction, a small decoder head (five $1{\times}1$ convolutions interleaved with bilinear $2\times$ upsampling) maps the $7{\times}7{\times}d$ feature map to a $224{\times}224$ binary mask; we train the decoder for 50 epochs with Adam (initial learning rate $10^{-3}$, cosine schedule), batch size 16, combined BCE + Dice loss, no augmentation, three seeds.
A key-holding party applies \texttt{spatial\_unpermute} before decoding; this step has no effect on classification since global average pooling is permutation-invariant, but is required for dense prediction.
To verify that key-holding users can still recover spatial explanations despite the permutation, we additionally train a spatial-attention classifier on PneumoniaMNIST: a linear layer on the $7 \times 7$ feature map without global average pooling, followed by CAM extraction via the learned spatial weights, which the authorized user inverse-permutes back to the original coordinate system.
We measure preservation by Pearson correlation between CAMs from original and \method{}-transformed features.

\paragraph{Privacy metrics.}
We evaluate privacy through three complementary metrics:
(1)~\emph{Reconstruction SSIM}: a U-Net decoder (5 upsampling blocks, $7{\times}7 \to 224{\times}224$, trained on 5000 samples with MSE loss for 25 epochs) reconstructs the original image from features; higher SSIM indicates more visual information leakage.
(2)~\emph{Verification AUC}~\citep{packhaeuser2022reid}: a cosine-similarity-based protocol tests whether features from the same patient are more similar than features from different patients, using 1000 balanced pairs (500 same-patient, 500 different-patient) drawn from 500 held-out patients; AUC~0.5 indicates random chance.
(3)~\emph{Top-1 accuracy}: the fraction of queries where the closest feature match in a gallery of 500 patients is the correct patient.

\paragraph{Baselines and ablations.}
We compare \method{} against eight methods including baselines and ablations (all at $d{=}512$ unless noted):
(1)~Raw features (no defense),
(2)~Gaussian noise ($\sigma = 3.0$),
(3)~Spatial permutation only,
(4)~\emph{\method{} w/o permutation} (ReLU + key),
(5)~\emph{\method{} w/o key} (non-secret $\mathbf{W}$),
(6)~\emph{\method{} w/o ReLU} (linear, keyed),
(7)~\method{} (full pipeline),
(8)~DP-Gaussian~\citep{dwork2014algorithmic}.
We also compare against NoPeek~\citep{vepakomma2020nopeek} applied post-hoc to frozen features; Shredder~\citep{mireshghallah2020shredder} and DISCO~\citep{singh2021disco} require training-time integration and cannot be applied post-hoc.
Our ablations subsume two prior keyed methods: \emph{Spatial Perm} corresponds to the patch shuffling approach of \citet{yao2022patchshuffling}, and \emph{KNT w/o ReLU} captures the same linear-invertibility vulnerability as the orthogonal mixing of PPCMI-SF~\citep{bello2026ppcmisf}, though PPCMI-SF operates on learned autoencoder latents rather than frozen classification features.
We now define the adversary under which these metrics are evaluated.

\subsection{Threat Model}
\label{sec:threat}

We consider an \emph{honest-but-curious} adversary: a cloud server that faithfully executes inference but attempts to extract patient identity from the received features.
We analyze two adversary regimes based on the attacker's knowledge of the secret key.

In the \emph{primary regime}, the adversary has access to the transmitted features and knows the feature extractor architecture (ResNet-18), but does not possess the secret key~$k$.
The server never sees the original images and is assumed not to have access to supervised training pairs for a learned re-identification model; its primary attack surface is unsupervised comparison of the transformed features it receives.
We use cosine-similarity-based verification as our primary privacy metric, as it is a standard and strong unsupervised matching baseline for this adversary.

In the \emph{key-compromise regime}, the adversary additionally obtains full knowledge of the key~$k$ and all derived parameters (weight matrices, biases, permutation).
This models insider threats, server breaches, or operational misconfiguration.
The adversary may attempt to recover original features via analytic inversion or gradient-based optimization.
We evaluate this regime separately (Section~\ref{sec:keycompromise}) to quantify how much protection the nonlinear transform provides independent of key secrecy.
Reconstruction attacks are a related threat; we evaluate reconstruction quality via SSIM.

Dense-prediction tasks introduce an additional authorized party: the decoder owner, who holds the key in order to inverse-permute the features before the task head (Section~\ref{sec:main}). This party is inside the trust boundary; our threat model applies to an external server that receives only the \method{}-transformed features.

\subsection{Main Results}
\label{sec:main}

\begin{figure}[!htbp]
  \centering
  \includegraphics[width=\linewidth]{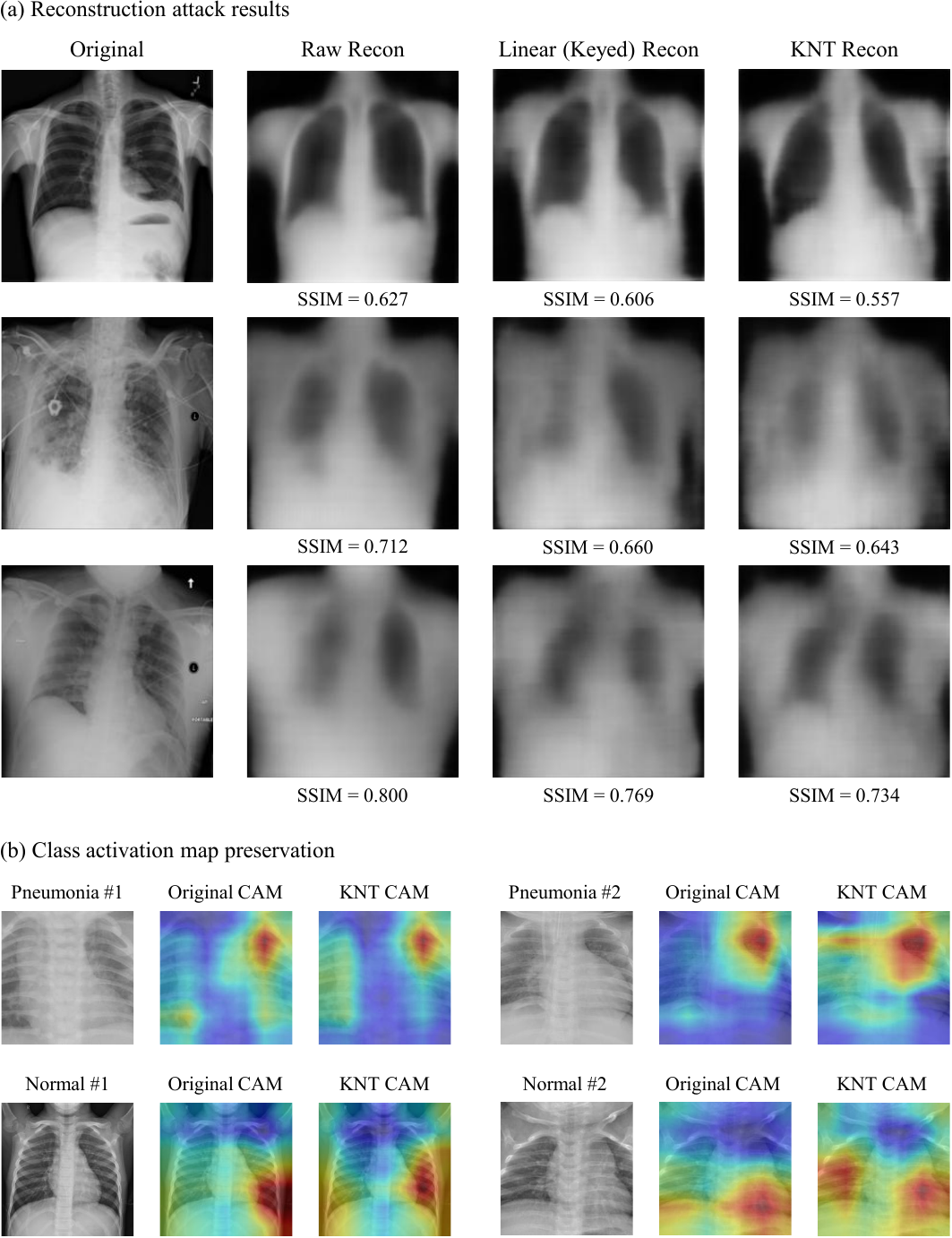}
  \caption{\textbf{Qualitative privacy and utility analysis.} \textbf{(A)}~Reconstruction comparison between raw, linear-keyed, and KNT features. \textbf{(B)}~CAM preservation on PneumoniaMNIST after inverse permutation (r = 0.673; without key: r = 0.025).}
  \label{fig:qualitative}
\end{figure}

\paragraph{Classification.}
Table~\ref{tab:main} presents the main comparison across all methods on ChestMNIST at $d{=}512$.
\method{} (full 2-layer pipeline) reduces identity signal the most: test-set verification AUC drops from 0.635 (raw features) to 0.586 (3 key seeds), and Top-1 re-identification drops from 1.3\% to 0.5\%.
The classification cost is modest: test AUC decreases from 0.759 to 0.749, a 1.0 percentage point drop.
Spatial permutation alone incurs zero classification cost (AUC 0.759, identical to raw) because global average pooling is permutation-invariant, yet already provides a meaningful privacy improvement (verification AUC 0.624).
In terms of above-chance identity signal, raw features score $0.635 - 0.5 = 0.135$ above random chance; \method{} reduces this to $0.586 - 0.5 = 0.086$, a 36\% relative reduction.
Figure~\ref{fig:qualitative}(a) visualizes the SSIM ordering: U-Net reconstructions from raw, linear keyed (no ReLU), and \method{} features, with per-image annotations matching the Table~\ref{tab:main} means (raw .722 $>$ linear .680 $>$ \method{} .665).
NoPeek~\citep{vepakomma2020nopeek} is designed for training-time integration; under the post-hoc frozen-feature conditions evaluated here (Table~\ref{tab:main}, $\dagger$), it achieves 0.723 classification AUC vs.\ \method{}'s 0.749, though this gap may partly reflect the adaptation mismatch.

\begin{table}[t]
  \centering
  \caption{\textbf{Main results on ChestMNIST ($d{=}512$, 2-layer).} All methods are applied post-hoc to frozen ResNet-18 layer-4 features. Classification AUC is the mean over 3 linear-probe seeds (std $\leq 0.002$, omitted for clarity). Privacy metrics use the $\pm$ notation for \emph{transformation seed} variation: 3 key seeds for keyed methods, 3 noise seeds for Gaussian noise. Methods without randomized transforms (Raw, \method{} w/o key) have no such variation.}
  \label{tab:main}
  \smallskip
  \resizebox{\linewidth}{!}{%
  \begin{tabular}{@{}lcccc@{}}
    \toprule
    \textbf{Method} & \textbf{Cls AUC}$\uparrow$ & \textbf{Recon SSIM}$\downarrow$ & \textbf{Verif AUC}$\downarrow$ & \textbf{Top-1}$\downarrow$ \\
    \midrule
    Raw features (no defense) & .759 & .722 & .635 & 1.3\% \\
    Gaussian noise ($\sigma{=}3.0$) & .722 & $.647 \pm .003$ & $.610 \pm .003$ & $0.7 \pm 0.2$\% \\
    Spatial permutation (keyed) & .759 & $.686 \pm .004$ & $.624 \pm .006$ & $0.8 \pm 0.2$\% \\
    \midrule
    \method{} w/o permutation (ReLU + key) & .749 & $.698 \pm .002$ & $.634 \pm .006$ & $1.0$\% \\
    \method{} w/o key (non-secret $\mathbf{W}$) & .747 & .664 & .590 & 0.4\% \\
    \method{} w/o ReLU (linear, keyed) & .754 & $.680 \pm .005$ & $.607 \pm .010$ & $0.6 \pm 0.1$\% \\
    \midrule
    \textbf{\method{} (full, 2-layer)} & \textbf{.749} & $\mathbf{.665} \pm .004$ & $\mathbf{.586} \pm .005$ & $\mathbf{0.5} \pm 0.1$\% \\
    \midrule
    NoPeek$^\dagger$~\citep{vepakomma2020nopeek} & .723 & .691 & .615 & 0.6\% \\
    \bottomrule
  \end{tabular}%
  }
  \vspace{1pt}
  {\footnotesize $^\dagger$Applied post-hoc to frozen features; the original method requires training-time integration.}
\end{table}

\paragraph{Localization.}
Beyond label prediction, we ask whether authorized users can recover spatially meaningful explanations after permutation.
On PneumoniaMNIST, the spatial-attention classifier described in Section~\ref{sec:setup} trained on \method{} features reaches AUC 0.980, matching the one trained on original features (0.981).
Figure~\ref{fig:qualitative}(b) shows CAM preservation: after inverse permutation with the authorized key, the correlation with the original-feature CAM is $r = 0.673$, meaning diagnostically relevant spatial attention patterns survive the transform.
Without the key, the correlation collapses to $r = 0.025$, confirming that spatial localization becomes substantially less accessible without the key.

\paragraph{Segmentation.}
To verify that \method{} extends to dense prediction, we evaluate pixel-level skin-lesion segmentation on ISIC 2018 Task~1 (Section~\ref{sec:setup}).
The backbone and \method{} transform are identical to the classification setup; only the task head differs, using a bilinear-upsample decoder in place of the linear probe.
Table~\ref{tab:seg} reports mean Dice and IoU on the held-out test set over three seeds.
At $d{=}512$, \method{} preserves Dice within 4.4\,pp of the undefended Raw baseline ($0.754 \to 0.710$), and degrades monotonically with $d$.
A decoder without access to the key, skipping the inverse permutation, achieves only Dice $0.542$ at $d{=}512$ (Table~\ref{tab:seg}, bottom row), a 21\,pp drop confirming that the decoder cannot implicitly learn to reverse the permutation during training.
Because the underlying transform is unchanged, feature-level privacy metrics reported in Section~\ref{sec:main} and Section~\ref{sec:keycompromise} transfer directly; output-space leakage through the predicted masks themselves is task-specific and left for future work.

\begin{table}[t]
  \centering
  \caption{\textbf{Dense prediction extension.} ISIC 2018 Task~1 skin lesion segmentation, held-out test set (1000 images), 3 seeds. $\Delta$: absolute Dice drop from Raw. Decoder trained on top of frozen ResNet-18 layer-4 features with the authorized-use protocol (inverse spatial permutation before decoding). $^\dagger$Decoder trained without inverse permutation, simulating an unauthorized party.}
  \label{tab:seg}
  \smallskip
  \begin{tabular}{@{}lccc@{}}
    \toprule
    \textbf{Method} & \textbf{$d$} & \textbf{Dice} & \textbf{IoU} \\
    \midrule
    Raw features           & 512 & $.754 \pm .002$             & $.630 \pm .002$ \\
    \method{}              & 512 & $.710 \pm .005\ (-4.4)$     & $.577 \pm .004$ \\
    \method{}              & 256 & $.690 \pm .005\ (-6.4)$     & $.553 \pm .005$ \\
    \method{}              & 128 & $.676 \pm .003\ (-7.8)$     & $.537 \pm .002$ \\
    \midrule
    \method{} w/o unpermute$^\dagger$ & 512 & $.542 \pm .006\ (-21.2)$  & $.371 \pm .005$ \\
    \bottomrule
  \end{tabular}
\end{table}

\subsection{Analysis}
\label{sec:analysis}

\paragraph{Component ablation.}
Table~\ref{tab:main} quantifies each component's contribution.
\emph{Spatial permutation} provides the largest single privacy improvement: verification AUC drops from 0.635 (raw) to 0.624, with zero classification cost.
Adding permutation to \emph{KNT w/o permutation} ($0.634 \to 0.586$) yields an additional 0.048 reduction, so spatial disruption and the nonlinear transform provide additive privacy gains.
The activation function has a modest empirical effect under the primary regime: at $d{=}512$ with 2 layers, the linear variant (verification AUC 0.607) is close to the nonlinear one (0.586).
However, linear transforms are analytically invertible given the key~\citep{shi2025scalemia}, whereas ReLU breaks closed-form inversion; the key-compromise regime exposes this gap (Section~\ref{sec:keycompromise}).
The \emph{KNT w/o key} ablation (weights derived from a fixed, non-secret seed, verification AUC 0.590) confirms that the cosine-similarity attack does not require key knowledge, since it compares features without attempting inversion.
The key becomes critical when an adversary attempts feature recovery, as analyzed in Section~\ref{sec:keycompromise}.

We sweep the number of ReLU layers from $L{=}1$ to $L{=}4$ (Table~\ref{tab:depth}).
Each additional layer improves privacy at increasing utility cost: $L{=}1 \to 2$ drops verification AUC from 0.595 to 0.586 at 0.6\,pp classification cost, while $L{=}3 \to 4$ drops verification by only 0.002 at 1.3\,pp classification cost.
We use $L{=}2$ as the default for its best privacy-per-utility-cost ratio.

\begin{table}[t]
  \centering
  \caption{\textbf{Layer depth ablation on ChestMNIST ($d{=}512$).} Classification AUC and privacy metrics vs.\ number of ReLU layers. Std omitted for clarity. Privacy gains diminish after $L{=}2$; utility cost accelerates.}
  \label{tab:depth}
  \smallskip
  \begin{tabular}{@{}cccc@{}}
    \toprule
    $L$ & \textbf{Cls AUC}$\uparrow$ & \textbf{SSIM}$\downarrow$ & \textbf{Verif AUC}$\downarrow$ \\
    \midrule
    1 & .755 & .672 & .595 \\
    \textbf{2} & \textbf{.749} & .665 & .586 \\
    3 & .743 & .649 & .577 \\
    4 & .730 & .633 & .575 \\
    \bottomrule
  \end{tabular}
\end{table}

\paragraph{Generalization across datasets, backbones, and dimensions.}
We evaluate \method{}'s generalization along three axes: projection dimension~$d$, dataset diversity, and backbone architecture.
Table~\ref{tab:dsweep} reports classification AUC across three datasets and three projection dimensions.
At $d = 512$, utility drops are at most 2.8\,pp across all three datasets evaluated with ResNet-18; at $d = 256$ ($2\times$ compression), drops remain within 6.2\,pp; at $d = 128$ ($4\times$ compression), ChestMNIST loses 4.2\,pp, while DermaMNIST and RetinaMNIST lose 11.4\,pp and 6.7\,pp respectively; fine-grained datasets are more sensitive to aggressive compression.
To confirm backbone agnosticism, we evaluate on ResNet-50 (Table~\ref{tab:dsweep}, bottom rows): original AUC 0.762, \method{} at $d{=}512$ yields 0.733 (a 3.0\,pp drop).
At $d{=}C{=}2048$ (matching the backbone output, no projection), the drop shrinks to 0.4\,pp (0.762 $\to$ 0.758), confirming that utility loss at lower $d$ is due to projection, not the nonlinear transform itself.

\begin{table}[t]
  \centering
  \caption{\textbf{Generalization across datasets and projection dimensions ($L{=}2$).} Classification AUC for \method{} at varying $d$ on ChestMNIST, DermaMNIST, and RetinaMNIST (ResNet-18) plus ChestMNIST (ResNet-50). $\Delta$: absolute drop from the undefended baseline. At $d{=}C$ (no projection), drops are minimal.}
  \label{tab:dsweep}
  \smallskip
  \resizebox{\linewidth}{!}{%
  \begin{tabular}{@{}lccccc@{}}
    \toprule
    \textbf{Dataset} & \textbf{Original} & \textbf{$d{=}128$} ($\Delta$) & \textbf{$d{=}256$} ($\Delta$) & \textbf{$d{=}512$} ($\Delta$) & \textbf{$d{=}C$} ($\Delta$) \\
    \midrule
    \multicolumn{6}{@{}l}{\textit{ResNet-18 ($C{=}512$)}} \\
    ChestMNIST (14-label) & .759 & .717 (${-}4.2$) & .736 (${-}2.3$) & \multicolumn{2}{c}{.749 (${-}1.0$)} \\
    DermaMNIST (7-class) & .925 & .811 (${-}11.4$) & .863 (${-}6.2$) & \multicolumn{2}{c}{.897 (${-}2.8$)} \\
    RetinaMNIST (5-class) & .841 & .774 (${-}6.7$) & .802 (${-}3.8$) & \multicolumn{2}{c}{.825 (${-}1.6$)} \\
    \midrule
    \multicolumn{6}{@{}l}{\textit{ResNet-50 ($C{=}2048$)}} \\
    ChestMNIST (14-label) & .762 & .689 (${-}7.3$) & .714 (${-}4.8$) & .733 (${-}3.0$) & .758 (${-}0.4$) \\
    \bottomrule
  \end{tabular}%
  }
\end{table}

Table~\ref{tab:generalization} provides additional cross-dataset evidence.
On PathMNIST (9-class histology), \method{} at $d{=}512$ incurs only a 0.2\,pp AUC drop ($0.989 \to 0.987$); on PneumoniaMNIST (binary), the drop is 1.5\,pp ($0.980 \to 0.965$).
All timing is measured on a single AMD Ryzen~7 9800X3D CPU (no GPU): the \method{} transform runs in 0.15\,ms, with end-to-end latency 6.2\,ms (ResNet-18) and 17.0\,ms (ResNet-50).

\begin{table}[t]
  \centering
  \caption{\textbf{Cross-dataset generalization and efficiency.} \method{} ($d{=}512$, 2-layer) transfers across datasets with minimal utility loss and runs efficiently on CPU.}
  \label{tab:generalization}
  \smallskip
  \begin{tabular}{@{}lcc@{}}
    \toprule
    & \textbf{Original} & \textbf{\method{}} \\
    \midrule
    \multicolumn{3}{l}{\emph{Classification AUC by dataset ($d{=}512$, 2-layer)}} \\
    \quad ChestMNIST (14-label) & $.759 \pm .0002$ & $.749 \pm .0001$ \\
    \quad PathMNIST (9-class) & $.989 \pm .0002$ & $.987 \pm .0001$ \\
    \quad PneumoniaMNIST (binary) & $.980 \pm .0012$ & $.965 \pm .0015$ \\
    \midrule
    \multicolumn{3}{l}{\emph{Efficiency on CPU (AMD Ryzen 7 9800X3D, 64\,GB RAM)}} \\
    \quad ResNet-18 (+ \method{}) & 6.0\,ms & 6.2\,ms \\
    \quad ResNet-50 (+ \method{}) & 16.0\,ms & 17.0\,ms \\
    \bottomrule
  \end{tabular}
\end{table}

\begin{figure}[!htbp]
  \centering
  \includegraphics[width=0.7\linewidth]{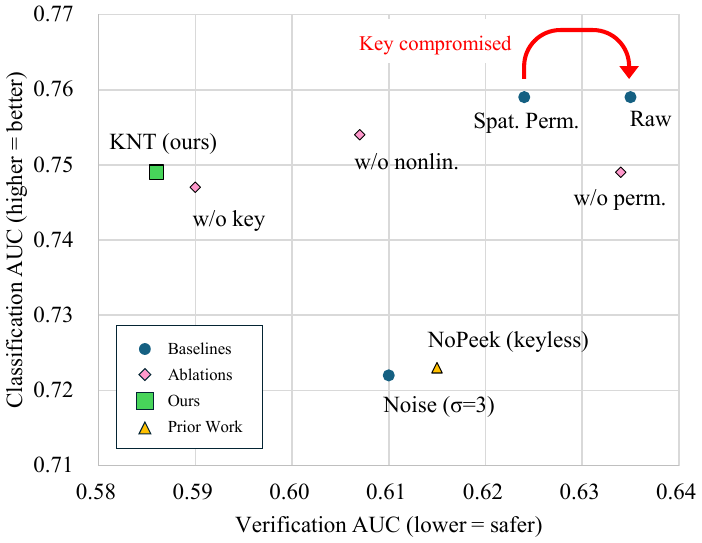}
  \caption{\textbf{Privacy-utility tradeoff on ChestMNIST.} KNT achieves the best empirical tradeoff in the upper-left region. Red arrow shows key compromise: spatial permutation degrades to raw. Gradient-based recovery under key compromise is analyzed in Section~\ref{sec:keycompromise}.}
  \label{fig:pareto}
\end{figure}

\paragraph{Privacy-utility tradeoff summary.}
Figure~\ref{fig:pareto} summarizes all methods in a privacy-utility scatter plot (classification AUC on $y$, verification AUC on $x$).
\method{} (full 2-layer, $d{=}512$) occupies the upper-left region with the lowest verification AUC (0.586) and only 1.0\,pp classification loss.
Gaussian noise ($\sigma = 3.0$) provides weaker privacy (verification AUC 0.610) at a higher classification cost (AUC 0.722); spatial permutation alone offers a zero-cost privacy improvement (AUC 0.759, verification 0.624).
However, the plot also exposes a weakness of spatial permutation: if the key is compromised, an attacker can reverse the permutation and recover raw-level privacy (verification AUC 0.635).
\method{} fares better: analytic inversion yields near-random features (cosine similarity .005), though gradient-based optimization can partially recover identity structure (Section~\ref{sec:keycompromise}).

\paragraph{Comparison with formal differential privacy.}
Table~\ref{tab:dp} compares \method{} against the Gaussian mechanism~\citep{dwork2014algorithmic} applied to $L_2$-clipped frozen features at $\delta = 10^{-5}$.
On $7{\times}7{\times}512 = 25{,}088$-dimensional layer-4 features, the calibrated noise destroys utility before reaching \method{}'s operating point: at $\varepsilon = 8$ (the conventional formal-privacy regime), classification AUC drops to 0.508; even at $\varepsilon = 128$, it recovers only to 0.591.
Under this frozen-feature setting, KNT achieves a more favorable empirical utility–identity-leakage tradeoff than the tested DP-Gaussian configurations, achieving both higher classification AUC (0.749 vs.\ 0.591) and lower verification AUC (0.586 vs.\ 0.519) than the most permissive budget.
The two defenses target complementary threats: \method{}'s keyed transform protects against feature interception by parties without the key, while DP protects against statistical inference of training-set membership.

\begin{table}[t]
  \centering
  \caption{\textbf{Comparison with formal Gaussian-mechanism DP~\citep{dwork2014algorithmic}.} Per-image $L_2$ clip at the 95th-percentile training-feature norm, then calibrated $\mathcal{N}(0, \sigma^2 I)$ at $\delta = 10^{-5}$. Three noise seeds $\times$ three classification seeds, test split. Raw baseline from Table~\ref{tab:main}.}
  \label{tab:dp}
  \smallskip
  \begin{tabular}{@{}lcccc@{}}
    \toprule
    \textbf{Method} & \textbf{$\sigma$} & \textbf{Cls AUC}$\uparrow$ & \textbf{Verif AUC}$\downarrow$ & \textbf{SSIM}$\downarrow$ \\
    \midrule
    Raw (Table~\ref{tab:main}) & 0 & .759 & .635 & .722 \\
    DP $\varepsilon{=}8$    & 364  & .508 & .497 & .477 \\
    DP $\varepsilon{=}128$  & 23   & .591 & .519 & .501 \\
    \midrule
    \textbf{\method{} (full)} & --- & \textbf{.749} & \textbf{.586} & \textbf{.665} \\
    \bottomrule
  \end{tabular}
\end{table}

\paragraph{Clinical-resolution validation.}
We replicate the main protocol on NIH ChestX-ray14~\citep{wang2017chestxray} at native $1024\times1024$ resolution (Table~\ref{tab:clinical}).
\method{} yields classification AUC 0.704 against a Raw baseline of 0.726 (2.2\,pp drop).
Verification AUC saturates near chance for both methods (0.536 vs.\ 0.537), so the binding privacy signal shifts to reconstruction: \method{} reduces SSIM from 0.838 to $0.672 \pm 0.004$, a 16.5\,pp gap, close to triple the 5.7\,pp gap at 224.

\begin{table}[t]
  \centering
  \caption{\textbf{Clinical-resolution validation on NIH ChestX-ray14~\citep{wang2017chestxray} ($1024\times1024$).} Frozen ImageNet ResNet-18, 3 cls seeds $\times$ 3 key seeds. Classification seed std $< 0.001$ (omitted). Verification AUC near chance for both methods; the privacy-relevant metric at this resolution is SSIM.}
  \label{tab:clinical}
  \smallskip
  \begin{tabular}{@{}lcccc@{}}
    \toprule
    \textbf{Method} & \textbf{Cls AUC}$\uparrow$ & \textbf{SSIM}$\downarrow$ & \textbf{Verif AUC}$\downarrow$ & \textbf{Top-1}$\downarrow$ \\
    \midrule
    Raw              & $.726$ & $.838$ & $.536$ & $0.02\%$ \\
    \method{}        & $.704$ & $.672 \pm .004$ & $.537 \pm .002$ & $0.05\%$ \\
    \bottomrule
  \end{tabular}
\end{table}

\subsection{Key-Compromise Security Analysis}
\label{sec:keycompromise}

We evaluate \method{}'s residual security when the key is fully compromised, as defined in Section~\ref{sec:threat}.

\paragraph{Analytic inversion.}
A linear \method{} (no ReLU) is trivially invertible: the pseudoinverse recovers original features exactly (cosine similarity 1.0, Top-1 100\%).
Applying the same pseudoinverse to the nonlinear variant yields near-random features (cosine similarity 0.005, Top-1 0.1\%), confirming that ReLU breaks closed-form inversion.

\paragraph{Gradient-based inversion.}
Since the pseudoinverse cannot exploit the piecewise-linear structure of ReLU, we evaluate gradient-based optimization~\citep{erdogan2022unsplit,shi2025scalemia} minimizing
$\|\method{}(\hat{\mathbf{f}}, k) - \mathbf{g}\|^2 + \lambda\|\hat{\mathbf{f}}\|^2$
via Adam (5 per-sample restarts, 2000 steps, $\lambda{=}10^{-4}$).
Across 3 key seeds and 1000 test samples each, this attack recovers features with cosine similarity 0.626 and Top-1 retrieval of 99.8\%, requiring approximately 15\,s per sample on an RTX 3090 Ti.
The recovered features preserve enough identity-discriminative structure for near-perfect re-identification, though they are not exact copies.

\paragraph{Effect of dimensionality reduction.}
Dimensionality reduction provides an additional defense even under full key compromise.
Reducing $d$ from 512 to 128, the gradient attack's cosine similarity drops from 0.626 to 0.271 and Top-1 retrieval falls from 99.8\% to 28.8\%, because the $512 \to 128$ projection permanently discards information that the optimizer cannot recover.
Table~\ref{tab:keycompromise} summarizes these results.

\begin{table}[t]
  \centering
  \caption{\textbf{Key-compromise security analysis.} Gradient-based inversion results (3 key seeds, 1000 test samples each). Dimensionality reduction substantially degrades attack performance.}
  \label{tab:keycompromise}
  \smallskip
  \begin{tabular}{@{}lccc@{}}
    \toprule
    \textbf{Attack} & \textbf{$d$} & \textbf{Cosine Sim} & \textbf{Top-1} \\
    \midrule
    Pseudoinverse (linear) & 512 & $1.000$ & $100\%$ \\
    Pseudoinverse (nonlinear) & 512 & $.005$ & $0.1\%$ \\
    \midrule
    Gradient (nonlinear) & 512 & $.626 \pm .005$ & $99.8 \pm 0.2\%$ \\
    Gradient (nonlinear) & 256 & $.404 \pm .014$ & $71.9 \pm 5.5\%$ \\
    Gradient (nonlinear) & 128 & $.271 \pm .014$ & $28.8 \pm 3.0\%$ \\
    \bottomrule
  \end{tabular}
\end{table}

\paragraph{Implications.}
Nonlinearity shifts inversion from closed-form recovery to iterative optimization (${\sim}$15\,s per sample vs.\ instantaneous), but does not prevent feature recovery.
The recovered objects are approximations of the feature vectors, not the original images; faithful inversion from deep features remains extremely difficult~\citep{xu2024fora}.
The primary additional risk under key compromise is therefore cross-database re-identification; classification and attribute inference are already possible from the transformed features without the key~\citep{melis2019unintended}.
\method{}'s privacy therefore depends principally on key secrecy, with nonlinearity and dimensionality reduction providing secondary barriers.

\section{Discussion}
\label{sec:discussion}

\method{} is designed for institutions that need lightweight feature protection beyond unprotected split inference, but cannot afford federated learning or cryptographic protocols.
Compared to federated learning, which requires GPU infrastructure, synchronous training, and aggregation servers at every institution, \method{} requires only a frozen feature extractor with negligible per-image overhead (Table~\ref{tab:generalization}).
Homomorphic encryption and secure multi-party computation provide provable security but are orders of magnitude more expensive.
In resource-constrained hospitals, a low-overhead privacy layer that can actually be deployed may matter more than provable security that cannot.
KNT is not intended to replace cryptographic or formally private inference protocols, but to provide a lightweight practical protection layer for split inference.

\method{}'s structured keyed transform reduces identity signal more effectively than unstructured perturbation (Gaussian noise) while incurring a smaller classification penalty (Table~\ref{tab:main}).
Some identity signal survives, as expected: \citet{xiao2024formal} proved that any utility-preserving encoding necessarily leaks information about its input.
Under key compromise, nonlinearity raises the attack cost from closed-form to iterative optimization but does not prevent recovery (Section~\ref{sec:keycompromise}); key secrecy is therefore the principal defense.

For deployment, we recommend standard key management practices (hardware security modules, key rotation, access separation).
Dimensionality reduction ($d{=}128$) provides an additional defense layer when bandwidth constraints permit the utility tradeoff.
Per-institution keys require only per-key retraining of the server-side linear probe; the transform itself operates on any spatial feature map $H \times W \times C$ and is backbone-agnostic.

\paragraph{Limitations.}
Our re-identification evaluation assumes an honest-but-curious server without access to supervised identity-labeled training pairs; stronger adversaries with auxiliary labeled data could train learned re-identification models that outperform cosine similarity. Reconstruction leakage is evaluated separately with a supervised decoder (Section~\ref{sec:setup}).
At $1024\times1024$, the frozen ImageNet-pretrained backbone yields a lower baseline classification AUC than at $224\times224$ (0.726 vs.\ 0.759); fine-tuning or pretraining at native clinical resolution may recover this gap.
Detection benchmarks and extension to shallower feature layers remain future work.

\section{Conclusion}
\label{sec:conclusion}

Split inference provides a practical alternative to transmitting raw medical images, but intermediate features still leak patient identity information. KNT introduces lightweight key-conditioned obfuscation for split inference, reducing identity leakage and reconstruction quality while preserving downstream performance with negligible overhead and no backbone retraining. Our analysis shows that nonlinearity shifts inversion from closed-form recovery to iterative optimization, while dimensionality reduction provides additional robustness under key compromise. With proper key management, KNT offers a lightweight practical protection layer between unprotected feature sharing and computationally expensive cryptographic methods.

\paragraph{Reproducibility.}
Code and experiment configurations will be released upon publication.

\bibliography{references}

\end{document}